
\global\newcount\meqno
\def\eqn#1#2{\xdef #1{(\secsym\the\meqno)}
\global\advance\meqno by 1 $$#2\eqno#1$$}
%
\global\newcount\refno\def\ref#1#2#3{\global\advance\refno by1
\xdef #1{[\the\refno]}\xdef #2{#3}#1}
\def\underarrow#1{\mathrel{\mathop{\longrightarrow}\limits_{#1}}}
\def\svec#1{\skew{-2}\vec#1}
\def\Tr{\,{\rm Tr}\,}

\magnification=1200

\vsize=7.5in
\hsize=5.6in

\baselineskip 12pt plus 1pt minus 1pt
\centerline{{\bf THE NUCLEON ``TENSOR CHARGES'' AND THE SKYRME MODEL}
\footnote{$^\star$}{This work is supported in part by funds provided by the
U.S. Department of Energy (D.O.E.) under contract
\#DE-AC02-76ER03069.\smallskip}}
\vskip 24pt
\centerline{James M. Olness \footnote{$^\dagger$}{National Science
Foundation Graduate Fellow}}
\bigskip
\centerline{\it Center for Theoretical Physics}
\centerline{\it Laboratory for Nuclear Science}
\centerline{\it  and Department of Physics}
\centerline{\it Massachusetts Institute of Technology}
\centerline{\it Cambridge, MA 02139}
\vskip 1.5in
\centerline{Submitted to: \ Physical Review D (Brief Reports)}
\vfill
\noindent CTP\#2122\hfill July 1992
\eject
\centerline{\bf ABSTRACT}
\bigskip
The lowest moment of the twist-two, chiral-odd parton
distribution $h_1(x)$ of the nucleon can be related to the so-called
``tensor charges'' of the nucleon.  We consider the tensor charges in
the Skyrme model, and find that in the large-$N_c$, SU(3)-symmetric
limit, the model predicts that the octet isosinglet tensor charge, $g^8_T$,
is of order $1/N_c$ with respect to the octet isovector
tensor charge, $g^3_T$.  The predicted $F/D$ ratio is then 1/3,
in the large-$N_c$ limit.  These predictions coincide with the Skyrme
model predictions for the octet ${\it axial}$ charges,
$g^8_A$ and $g^3_A$.  (The prediction $F/D=1/3$ for the axial charges
differs from the commonly quoted prediction of 5/9, which is based on
an inconsistent treatment of the large-$N_c$ limit.)
The model also predicts that
the singlet tensor charge, $g^0_T$, is
of order $1/N_c$ with respect to $g^3_T$.
\vfill
\eject
\xdef\secsym{ }\global\meqno = 1
\medskip
The nucleon has three parton distributions at lowest twist,
that is, at twist-two.  Two of these, $f_1(x)$ and $g_1(x)$, have been studied
extensively, and have been measured in deep-inelastic scattering experiments
\ref\f\fref{For a review and references, see A. Manohar,
UCSD/PTH 92-10 (1992).}.
The remaining distribution, $h_1(x)$, is relatively new, and has only recently
begun to receive attention in the literature
\ref\h\href{For a review and references, see R. L. Jaffe and Xiangdong Ji,
{\it Nucl. Phys.\/} {\bf B375} (1992) 527.}.
It has not been measured $-$ since it is
chiral-odd, it is inaccessible to inclusive deep-inelastic scattering
experiments.  Ralston and Soper
\ref\ral\ralref{J. Ralston and D. E. Soper, {\it Nucl. Phys.\/} {\bf B152}
(1979) 109.}
have shown, however, that $h_1(x)$ plays an important role in polarized
Drell-Yan processes.  More recently, Collins
\ref\cll\cllref{J. Collins, Penn State Preprint, 1990.}
showed that it emerges naturally in the factorization of a general hard
process into soft and hard sub-processes.  In the parton model language,
$h_1(x)$ can be interpreted loosely as counting transversely polarized
(valence)
quarks in a transversely polarized nucleon.\footnote{$^\star$}
{A more precise interpretation involves the transversely projected
Pauli-Lubanski operator; see reference \h.}
The $Q^2$ evolution of $h_1(x)$ has been calculated by Kodaira {\it et al.\/}
\ref\kod\kodref{J. Kodaira, S. Matsuda, K. Sasaki, and T. Uematsu,
{\it Nucl. Phys.\/} {\bf B159} (1979) 99.},
and by Artru and Mekhfi
\ref\art\artref{X. Artru and M. Mekhfi, {\it Z. Physik\/}
{\bf C45} (1990) 669.}.

Sum rules relating the moments of $h_1(x)$ to the nucleon matrix elements of
local operators can be formulated in the standard way.  The low moments are
particularly interesting because they can be used to determine the gross
features of the distribution.  The lowest moment, for example, is referred
to as the ``tensor charge,'' $\delta q$ :
\eqn\deltaq{\delta q(Q^2)=\int^\infty_{-\infty} dx\,h_1(x,Q^2)=
\int^1_0 dx\,\Bigl(h_1(x,Q^2)-{\overline h_1}(x,Q^2)\Bigr)\ \ .}
Note that $h_1(x)$ carries flavor indices, which have been suppressed; there
is an independent tensor charge, $\delta q$, for each quark flavor.  The
tensor charge derives its name from its relation to the following nucleon
matrix element:
\eqn\tensop{\langle PS|\ {\overline q}\sigma_{\mu\nu} i\gamma^5 q\bigg|
_{Q^2}\ |PS \rangle = {2 \over M}(S_\mu P_\nu - S_\nu P_\mu)\delta q(Q^2)}
where $q=u,d,s,\, etc.$  (The parameter $Q^2$ appearing in equations (1)
and (2) is a renormalization scale label, necessary to render $h_1(x)$
and the tensor charges well-defined in perturbative QCD.
The tensor charges do not mix with other operators under renormalization,
and are therefore characterized by a single anomalous dimension, which
has been calculated at one loop [5,6]: $\gamma = 2\alpha_s / 3\pi$.
Henceforth, we will suppress all dependence on $Q^2$.)
It is convenient to arrange the $\delta q$'s into octet and singlet
combinations, having definite flavor SU(3) transformation properties:
$$\eqalign{g^3_T & = 2 (\delta u - \delta d) \cr
g^8_T & = {2 \over {\sqrt 3}} (\delta u + \delta d - 2\delta s) \cr
g^0_T & = 2{\sqrt {2 \over 3}} (\delta u + \delta d + \delta s)\ \
.\cr }\eqno(3)$$
The reason for the peculiar normalization will become clear shortly.
For future reference, we also define the axial charges, $g^a_A$:
$$\langle PS|{\overline \Psi} \lambda^a \gamma_\mu
\gamma^5 \Psi|PS \rangle = S_\mu g^a_A\ \ . \eqno(4)$$
This is just the conventional definition of the axial charges.
In the non-relativistic quark model, the tensor charges are equal
to the corresponding axial charges.  (We refer here to the octet
and singlet combinations, $g^a_T$ and $g^a_A$, {\it not} the individual
quark components, $\delta q$, and their axial charge counterparts,
$\Delta q$.  According to the conventional definition of $\Delta q$,
{\it i.e.\/}, $\langle PS|{\overline q} \gamma_\mu
\gamma^5 q|PS \rangle = S_\mu \Delta q$, we find that $\Delta q = 2\delta q$,
in the non-relativistic quark model.  In this paper, we will work
almost exclusively with $g^a_T$ and $g^a_A$.)
Jaffe and Ji \h \  considered the tensor charges in the bag model,
and found the tensor charges to be somewhat larger in absolute magnitude
than the corresponding axial charges.
It should be remembered, however, that this comparison involves
an implicit, unknown scale, of the order of 1 GeV ({\it i.e.\/},
 the ``bag'' scale
at which the operators are renormalized).

The purpose of this note is to report the results of a Skyrme model
analysis of the tensor charges.
\footnote{$^\star$}{We work with the Skyrme
model in order to be concrete.  However, the conclusions are
valid in the large-$N_c$ limit of any chiral soliton model
in which the soliton is quantized semi-classically, through the use of
collective coordinates.}
In the large-$N_c$, SU(3)-symmetric limit,
the model predicts that the octet isosinglet tensor charge, $g^8_T$, is of
order $1/N_c$ with respect to the (octet) isovector
tensor charge, $g^3_T$:
$${g^8_T \over g^3_T} = O\left({1 \over N_c}\right)\ \ . \eqno(5)$$
This can also be phrased in terms of a prediction for the $F/D$ ratio:
$${F \over D} = {1 \over 3} + O\left({1 \over N_c}\right)\ \ . \eqno(6)$$
These predictions coincide with the large-$N_c$, SU(3)-symmetric Skyrme
model predictions for the octet ${\it axial}$ charges, $g^8_A$ and $g^3_A$.
(The prediction $F/D=1/3$ for the axial charges differs from the commonly
quoted prediction of 5/9, which is based on an inconsistent treatment of
the large-$N_c$ limit.)
The model also predicts that the singlet tensor charge, $g^0_T$, is of
order $1/N_c$ with respect to $g^3_T$:
$${g^0_T \over g^3_T} = O\left({1 \over N_c}\right)\ \ . \eqno(7)$$
(In the case of the ${\it axial}$ charges, the model
predicts $g^0_A/g^3_A=O\bigl({1 \over N^2_c}\bigr)$
\ref\bek\bekref{S. Brodsky, J. Ellis, and M. Karliner,
{\it Phys. Lett.\/} {\bf B206} (1988) 309.}.)
These predictions are independent of the renormalization scale in
QCD, since the tensor charges have a common anomalous dimension.

Before considering the Skyrme model, it will be useful to discuss
the non-relativistic quark model (NQM).  As mentioned above, the tensor
charges are equal to the corresponding axial charges in the NQM.  This
is true for any value of $N_c$.  The explicit calculation of the axial
charges (and thus, the tensor charges) for arbitrary $N_c$ has been
carried out by Karl and Paton
\ref\kar\karref{G. Karl and J. E. Paton,
{\it Phys. Rev.\/} {\bf D30} (1984) 238.},
using the fact that the axial charges measure the spin carried by the
quarks.  Transcribing their results, we have
$$\eqalign{ g^3_T & = {1 \over 3} (N_c + 2) \cr
            g^0_T & = {\sqrt 2}\, g^8_T = {\sqrt {2 \over 3}}\ \
. \cr} \eqno(8)$$
In particular, the NQM predicts that $g^8_T/g^3_T$ and $g^0_T/g^3_T$ are
of order $1/N_c$.  The known correspondence
\ref\man\manref{A. Manohar, {\it Nucl. Phys.\/} {\bf B248} (1984) 19.}
between the large-$N_c$ NQM and the large-$N_c$ Skyrme model suggests that we
should expect similar results in the Skyrme model.

The Skyrme model
\ref\sky\skyref{E. Witten, {\it Nucl. Phys.\/} {\bf B223} (1983) 422, 433;
G. Adkins, C. Nappi, and E. Witten, {\it Nucl. Phys.\/} {\bf B228} (1983) 433;
E. Guadagnini, {\it Nucl. Phys.\/} {\bf B236} (1984) 35.}
describes the interactions of an octet of fundamental meson fields.
In its simplest form, the effective action is given by
$$\Gamma=
\int d^4 x \biggl({F^2_\pi \over 16}
\Tr(\partial_\mu U \partial^\mu U^{\dagger})
+{1 \over 32e^2}
\Tr[(\partial_\mu U)U^{\dagger},(\partial_\nu U)U^{\dagger}]^2 \biggr) +
N_c \Gamma_{WZ}\ \ , \eqno(9)$$
where $U(x)={\rm exp}\bigl(2i\lambda^a\phi^a(x)/F_\pi\bigr)$, and the
$\phi_a$'s are meson fields.  $F_\pi$ is the pion decay constant, and $e$ is a
phenomenological parameter.  $\Gamma_{WZ}$ is the so-called Wess-Zumino term,
which incorporates the effects of anomalies.  As is well known, the
classical field equations associated with this lagrangian admit stable
soliton solutions, the prototypical example of which is the hedgehog
solution:
$$U_0({\svec x}) = \left( \matrix{
{\rm exp}\bigl(iF(r){\svec \tau} \cdot \hat x \bigr) & 0 \cr 0 & 1 \cr} \right)
\eqno(10)$$
This is quantized by introducing collective coordinates to describe
the orientation of the soliton in the group space ${\rm SU}(3)_{\rm flavor}$
(and thus, simultaneously, ${\rm SU}(2)_{\rm spin}$):
$$U({\svec x},t)=A(t)U_0({\svec x})A^{-1}(t)\ \ , \eqno(11)$$
where $A(t)$ is a
time-dependent SU(3) matrix.  The resulting quantum states have the
quantum numbers of baryons.  By making a suitable choice of the
parameters $F_\pi$ and $e$, the spectrum of low-lying baryons can
be modeled approximately.

In order to apply the Skyrme model in estimating nucleon matrix elements,
one must first identify the appropriate Skyrme model operators, constructed
in terms of meson fields.  The tensor charges are not associated with any
symmetry of the QCD lagrangian, so their Skyrme model analogs cannot be
generated by means of the standard Noether's theorem route.
Other methods of constructing equivalent operators in the Skyrme model were
considered, such as the method of Goldstone and Wilczek
\ref\gol\golref{J. Goldstone and F. Wilczek, {\it Phys. Rev. Lett.\/}
{\bf 47} (1981) 986.},
but these apparently lead to ambiguous results.  At leading order in the
1/$N_c$ expansion, however, the analysis simplifies considerably,
and we can learn about the flavor structure of the
tensor charges simply by considering the transformation properties of the
operators appearing in equation (2).

Consider the following matrix element:
$$\langle PS|{\overline \Psi} \lambda^a \sigma_{\mu \nu}i\gamma^5 \Psi| PS
  \rangle = {1 \over M} (S_\mu P_\nu - S_\nu P_\mu) g^a_T\ \ . \eqno(12)$$
For a nucleon at rest, the tensor operator appearing in this equation
yields non-vanishing matrix elements only
for $(\mu ,\nu ) = (i,0), \, i=1,2,3$.  Denote these components of the
operator by $O^a_i$, and denote the equivalent Skyrme model
operators by $\widetilde{O}^a_i$.  The $\widetilde{O}^a_i$ transform
as a three-vector under spatial rotations, and as an octet under SU(3)
transformations (or as an SU(3) singlet, in the case of $a=0$).
In the large-$N_c$ limit, the soliton rotates
slowly, so as a first approximation, we ignore operators that involve
time-derivatives, or in other words, factors of the canonical momenta.
(The large-$N_c$ approximation
is standard in applications of the Skyrme model; indeed, it is assumed
implicitly in the quantization of the hedgehog solution.)  Then in direct
analogy with the quantum mechanics of a non-relativistic point particle,
we can evaluate the matrix elements of the
$\widetilde{O}^a_i$ in terms of SU(3) Clebsch-Gordan coefficients,
and a single (unknown) constant.  Standard Skyrme model formulas \man \ give
the following results for the octet charges:
$$\langle N | \widetilde{O}^a_m | N \rangle =
k\ \sum_{n=1,2} \left( \matrix {{\bf R} & {\bf 8} \cr N & a \cr} \ \bigg|
\ \matrix {{\bf R}_n \cr N } \right)
\left( \matrix {{\bf R} & {\bf 8} \cr N & b \cr} \ \bigg|
\ \matrix {{\bf R}_n \cr N } \right) \eqno(13)$$
The label $N$ represents a nucleon state, which we take to have definite
polarization in the $z$-direction, for simplicity.
The index $m$ indicates the $J_3$ spin quantum number of
the operator $\widetilde{O}^a$.  The label ${\bf R}$ denotes the SU(3)
representation containing the nucleon $-$ we will say more about
this shortly.  The direct product ${\bf R} \times {\bf 8}$ contains
two copies of the representation ${\bf R}$, which are denoted
by ${\bf R}_n$ in equation (13).
The label $b$ represents a state with hypercharge zero, and
isospin quantum numbers that are determined
by the transformation properties of $\widetilde{O}^a_m$ under spatial
rotations: $I=J=1$ and $I_3=-J_3=-m$.

The constant $k$ is unknown, but the scale-dependence of the tensor
charges in QCD means that precise knowledge of $k$ would be of questionable
value, since we do not know how to determine the renormalization
scale at which $k$ applies.  Note that $k$ may depend on $N_c$.
Indeed, as we will see, correspondence with the NQM suggests
that $k=O(N_c)$.

The corresponding formula for the singlet charge yields a vanishing
result, since there is no state $b$ having $I=1$ in the singlet
representation.  It should be emphasized that equation (13) and its
singlet counterpart
are valid only when we neglect the contributions of operators involving
time-derivatives.  The corrections that result from such operators
are discussed below.  For now, we point out that these corrections
do not necessarily vanish as $N_c$ becomes large.

We now discuss the identification of the large-$N_c$ analog of the
nucleon.  In a symmetric quark model, a baryon containing $N_c=2n+1$
quarks must belong to the (1,$n$) representation of SU(3) if it is to
have spin ${1 \over 2}$.  (The (1,$n$) representation consists of the
traceless, symmetric tensors having one upper index and $n$ lower
indices.)  Furthermore, for $N_c=2n+1$, the (1,$n$) representation
is the only representation which can be projected out of the soliton
solution and which contains states with zero strangeness
and $I=J={1 \over 2}$ \man.  A consistent
treatment of the large-$N_c$ limit therefore seems to require that
we choose ${\bf R} \, = \, (1,n)$ in equation (13).  (For a more
extensive discussion on this issue, see reference
\ref\kap\kapref{D. Kaplan and I. Klebanov,
{\it Nucl. Phys.\/} {\bf B335} (1990) 45.}.)
With this choice, equation (13) yields the following results:
$$\eqalign{g^3_T &=
     2k {(N_c-1)(N_c+6)^2+9(N_c+7) \over 3(N_c+7)(N_c+3)^2}
     \ \underarrow{N_c \to \infty} \ {2k \over 3} \cr
           g^8_T &=
     2k {2{\sqrt 3} \over (N_c+3)(N_c+7)}
     \ \underarrow{N_c \to \infty} \ {4{\sqrt 3}k \over N^2_c} \ \
.}\eqno(14)$$
Thus, the ratio $g^8_T/g^3_T$ vanishes in the large-$N_c$ limit,
although we will see that it is expected to vanish linearly in 1/$N_c$,
rather than quadratically. For comparison with the axial charges,
it is useful to translate this result into a statement about the
$F/D$ ratio.  According to the standard definition of the reduced
matrix elements, $F$ and $D$, the isosinglet tensor charge is given by:
$$g^8_T \sim {1 \over {\sqrt 3}} (3F - D)\ \ . \eqno(15)$$
Thus, the large-$N_c$ prediction is F/D = 1/3.  As mentioned previously,
this prediction is independent of the renormalization scale.

We now consider corrections to these results within the framework of the
Skyrme model.  Equation (13) is valid only if the operators that describe
the tensor charges do not involve time-derivatives of the collective
coordinates.  Presumably, the quark operators appearing in equation (12)
are matched onto an entire family of Skyrme model operators, and in
general, this family will include operators with time-derivatives.  We
can argue on dimensional grounds, however, that the contributions of these
operators are suppressed in the large-$N_c$ limit.  In the process of
reducing meson operators to collective coordinate operators
(by integrating over all space),
a space-derivative brings in a mass parameter of the order of the inverse
proton radius.  In contrast, a time-derivative brings in a mass parameter of
the order of the rotational kinetic energy of the soliton.  Thus, we consider
the ratio
$${L^2 / 2I \over 1 / R} \sim {L^2 \over MR} \eqno(16)$$
where $I$ is the moment of inertia, and $L$ is the angular momentum of the
spinning soliton.  The nucleon mass grows as $N_c$, while the radius remains
constant in the large-$N_c$ limit
\ref\lnc\lncref{E. Witten, {\it Nucl. Phys.\/} {\bf B160} (1979) 57.}.
Thus, we conclude that corrections arising from operators with
time-derivatives should be suppressed by 1/$N_c$.

In order to apply these ideas to the tensor charges, it is useful to
consider the individual quark components, $\delta u, \delta d$, and $\delta s$.
(This can be done unambiguously, since the octet and singlet tensor charges
are characterized by the same anomalous dimension.)
As suggested by the NQM, we will assume that $\delta u$ and $\delta d$ are
of order $N_c$.  Equation (14) then requires that $\delta s$ be at most
of order $N_c$.  Recalling the considerations of the preceding paragraph,
we find that operators with time-derivatives are expected to contribute
to the $\delta q$'s at order unity.  Equation (13) and its singlet
counterpart then imply that $g^3_T$ is of order $N_c$, while $g^0_T$
and $g^8_T$ are of order unity (and thus, finite, in the large-$N_c$ limit).
These conclusions,
of course, depend on the validity of our assumption that $\delta u$ and
$\delta d$ are of order $N_c$.  However, the conclusion that
$g^8_T/g^3_T$ and $g^0_T/g^3_T$ are of order $1/N_c$
is readily seen to be independent of this assumption, and can be taken
as a legitimate prediction of the Skyrme model.  From
$g^8_T/g^3_T=O\bigl({1 \over N_c}\bigr)$, we also obtain
$F/D=1/3+O\bigl({1 \over N_c}\bigr)$.

Now consider the axial charges.
For a nucleon at rest, only the space-components of
the axial-vector current operator appearing in equation (4)
yield non-vanishing matrix elements.  If we denote the space-components
by $A^a_i$, then the $A^a_i$ transform in exactly the same way as
the $O^a_i$ under ${\rm SU}(3)_{\rm flavor} \times {\rm SU}(2)_{\rm spin}$.
Thus, the analysis outlined above also applies to the axial charges,
and we draw the same conclusions as before:
The ratio $g^8_A/g^3_A$ is of order $1/N_c$,
and $F/D=1/3+O\bigl({1 \over N_c}\bigr)$; the ratio $g^0_A/g^3_A$ vanishes
in the large-$N_c$ limit, although, as mentioned previously, this ratio
is actually expected to vanish like $(1/N_c)^2$, rather than $1/N_c$.
(This last result does not contradict the arguments above, which are
only meant to place crude limits on the large-$N_c$ behavior.)

Note that the prediction $F/D=1/3$ for the axial charges differs from
the commonly quoted prediction of 5/9 (see, for example, reference
\ref\bij\bijref{J. Bijnens, H. Sonoda, and M. Wise,
{\it Phys. Lett.\/} {\bf B140} (1984) 421.}).
The latter prediction results
from choosing {\bf R}={\bf 8} in equation (13).  Although this value
({\it i.e.\/}, 5/9) is in good agreement
with the experimental result of $0.58 \pm 0.05$
\ref\jaf\jafref{R. L. Jaffe and A. Manohar,
{\it Nucl. Phys.\/} {\bf B337} (1990) 509.},
it is based on an
inconsistent treatment of the large-$N_c$ limit, so the agreement
should perhaps be regarded as fortuitous.  The proper prediction,
$F/D=1/3 + O\bigl({1 \over N_c}\bigr)$, clearly is not very precise
for the realistic case of $N_c=3$,
although it is at least consistent with the experimental result.
In the case of the tensor charges, where no experimental information
is available, it would seem that the only reasonable prediction
we can make is $F/D=1/3+O\bigl({1 \over N_c}\bigr)$.

The experimental result for the ratio $g^0_A/g^3_A$ is in fact small.
Using the experimental values $g^0_A=0.098 \pm 0.076 \pm 0.113$
and $g^3_A=1.254 \pm 0.006$
\ref\ash\ashref{J. Ashman {\it et al.\/}, {\it Nucl. Phys.\/} {\bf B328} (1989)
1.},
we obtain $g^0_A/g^3_A=0.078 \pm 0.109$.  Unfortunately, this result does
not offer a useful standard against which to compare the prediction
$g^0_T/g^3_T=O\bigl({1 \over N_c}\bigr)$, since $g^0_A/g^3_A$ is
expected to be of order $O\bigl({1 \over N^2_c}\bigr)$.

In the above analysis, we stated that $\delta s$ is at most of
order unity with respect to $\delta u$ and $\delta d$.
In fact, since $g^8_T$ and $g^0_T$ are both of
order $1/N_c$ with respect to $g^3_T$, $\delta s$ is
actually of order $1/N_c$ with respect to $\delta u$
and $\delta d$.  (It also follows that
$\delta u + \delta d$ is of order $1/N_c$ with respect to
$\delta u$ and $\delta d$, as in the NQM.)
The same conclusion can be drawn in the case of the
axial charges.  If $\Delta q$ is the axial charge counterpart
of the tensor charge $\delta q$, then $\Delta s$ is of
order $1/N_c$ with respect to $\Delta u$ and $\Delta d$.
Since $\Delta u$ and $\Delta d$ are of order $N_c$ \sky \ in the Skyrme
model, $\Delta s$, the fraction of the nucleon spin carried by strange
quarks, is of order unity, and not of order $N_c$, as previously
reported \bek.  This is consistent with the suppression of the $q{\overline q}$
sea that is expected on general grounds in large-$N_c$ QCD \lnc.  (The previous
prediction that $\Delta s = O(N_c)$ in the Skyrme model was based on an
``$N_c$-independent'' $F/D$ ratio of 5/9.  The argument used in that case
breaks down for $F/D=1/3$.)

Finally, we discuss the effects of SU(3) symmetry violation on the above
results.  The role of the strange quark in the nucleon is poorly understood
in general.  In the particular case of the tensor charges, it does not
seem possible to say anything rigorous about the effects of a
symmetry-breaking mass term for the strange quark.
Within the framework of the Skyrme model, for example, we might
approach this problem from the point of view of dimensional analysis.
Unfortunately, dimensional analysis appears to be completely unreliable
for SU(3) symmetry-breaking effects in the Skyrmion, since the
dimensionful parameter that sets the scale for symmetry-breaking
effects is of order 250 MeV ({\it i.e.\/}, less than the kaon mass, a typical
measure of symmetry violation in the chiral lagrangian) \kap.
We simply cannot
rule out the possibility of large corrections stemming from SU(3) violation.
For what it is worth, we point out that SU(3) appears to be a very good
symmetry in the case of the axial charges, although again, the reasons
for this are not well understood.
\bigbreak
\noindent{\bf SUMMARY}
\medskip
We consider the nucleon tensor charges in the Skyrme model.  We show that
in the large-$N_c$, SU(3)-symmetric limit, the model predicts
that $g^8_T/g^3_T$ is of order $1/N_c$.  This is
equivalent to the prediction
$F/D=1/3 + O\bigl({1 \over N_c}\bigr)$, which, unfortunately,
is not very precise for the realistic case of $N_c=3$.
These predictions are identical to the Skyrme model predictions for the
octet $axial$ charges.  The model also predicts that $g^0_T/g^3_T$ is
of order $1/N_c$.  All of these predictions are in
agreement with the predictions of the NQM.  Unlike the NQM, the Skyrme
model does not readily indicate the large-$N_c$ behavior of, say, $g^3_T$,
although it is not inconsistent to assume that $g^3_T$ is of order $N_c$
in the Skyrme model.  We point out that the commonly quoted prediction
for the axial charge $F/D$ ratio ({\it i.e.\/}, 5/9) is based on an
inconsistent
treatment of the large-$N_c$ limit.  A consistent treatment yields the
prediction quoted above, and shows that the strange quark contribution
to the nucleon spin is of order unity in the large-$N_c$ Skyrme model.
\bigbreak
\centerline{\bf ACKNOWLEDGEMENTS}
\bigbreak
I thank Robert Jaffe for suggesting this work, and for many
useful discussions.  I would also like to thank Matthias Burkhardt
and Eric Sather for useful discussions.
\bigbreak
\centerline{\bf REFERENCES}
\bigbreak
\item{\f}\fref
\medskip
\item{\h}\href
\medskip
\item{\ral}\ralref
\medskip
\item{\cll}\cllref
\medskip
\item{\kod}\kodref
\medskip
\item{\art}\artref
\medskip
\item{\bek}\bekref
\medskip
\item{\kar}\karref
\medskip
\item{\man}\manref
\medskip
\item{\sky}\skyref
\medskip
\item{\gol}\golref
\medskip
\item{\kap}\kapref
\medskip
\item{\lnc}\lncref
\medskip
\item{\bij}\bijref
\medskip
\item{\jaf}\jafref
\medskip
\item{\ash}\ashref
\medskip
\vfill
\eject
\end